\documentclass[pre, reprint, superscriptaddress, showpacs, showkeys, amsmath, amssymb, a4paper]{revtex4-1} 
\usepackage{graphicx,dcolumn,bm}

\begin{document}

\title{Analyses of kinetic glass transition in short-range attractive colloids \\based on time-convolutionless mode-coupling theory}

\author{Takayuki Narumi}
\email{narumi@ip.kyusan-u.ac.jp}
\affiliation{Faculty of Engineering, Kyushu Sangyo University, Fukuoka 813-8503, Japan}
\author{Michio Tokuyama}
\affiliation{Institute of Multidisciplinary Research for Advanced Materials, Tohoku University, Sendai 980-8577, Japan}

%
\date{\today}

\begin{abstract} 
The kinetic glass transition in short-range attractive colloids is theoretically studied by time-convolutionless mode-coupling theory (TMCT).
By numerical calculations, TMCT is shown to recover all the remarkable features predicted by the mode-coupling theory for attractive colloids, namely the glass-liquid-glass reentrant, the glass-glass transition, and the higher-order singularities.
It is also demonstrated through the comparisons with the results of molecular dynamics for the binary attractive colloids that TMCT improves the critical values of the volume fraction.
In addition, a schematic model of three control parameters is investigated analytically.
It is thus confirmed that TMCT can describe the glass-glass transition and higher-order singularities even in such a schematic model.
\end{abstract}
	


\maketitle

%%%%%%%%%%%%%%%%%%%%%%%%
\section{Introduction}
Short-range attractive colloids are prominent in studies of the glass transition.
In colloidal systems of a high volume fraction, since each particle is stuck in the ``cage'' made of the neighboring particles, the structural rearrangement rarely occurs.
The glass driven by the exclusive volume effect is classified as repulsive glass.
On the other hand, there is a different glass-forming mechanism in systems of attractive interaction.
At a low temperature, each particle is trapped in potential well and sticks together to form clusters.
The glass originated from the cluster formation is called attractive glass.
The attraction length in atomic or molecular systems is comparable to the particle size; nevertheless, in colloidal systems, the short-range attraction can be materialized \cite{Verduin1995, Mallamace2000, Segre2001, Eckert2002, Pham2002, Chen2002, Chen2003_science, Chen2003_PRE, Poon2004, Lu2008,  Eberle2012}.
For systems of the attraction range smaller than about one tenth of the particle diameter, the mode-coupling theory (MCT) predicted melting of a glass by cooling and the direct transition between repulsive and attractive glasses \cite{Fabbian1999, *Fabbian1999_erratum, Bergenholtz1999_PRE, Dawson2000, Zaccarelli2001}. 
Eckert et al.~and Pham et al.~then observed the glass-liquid-glass reentrant \cite{Eckert2002, Pham2002}, and Chen et al.~confirmed the glass-glass transition in experiments \cite{Chen2003_science}.
Numerical simulations for attractive colloids have also supported such rich phenomena \cite{Puertas2002, Foffi2002, Zaccarelli2002, Puertas2003, Sciortino2003, Zaccarelli2004, Zaccarelli2009}.
We here study the glass transition of short-range attractive colloids to validate a theory recently proposed by Tokuyama, time-convolutionless mode-coupling theory (TMCT) \cite{Tokuyama2014, Tokuyama2015}.

A glassy state is ideally characterized by the presence of an arrested part in correlation functions \cite{Edwards1975}, and MCT describes the kinetic glass transition as a nonlinear bifurcation, so-called nonergodic transition \cite{Bengtzelius1984, Leutheusser1984, Gotze1991, Gotze2009}.
However, while some extensions and modifications have been done \cite{Kawasaki2001, Szamel2003, Biroli2004, Biroli2006}, MCT has a few shortcomings that remain to be solved.
A fundamental problem is the case that the transition point predicted by MCT is far from the calorimetric glass transition points observed by experiments and also by simulations.
In order to overcome such a difficulty, TMCT has been proposed as an alternative theory of MCT \cite{Tokuyama2014}.

The way of extracting macroscopic (i.e., slow) dynamics differs between MCT and TMCT.
The starting equation of both MCT and TMCT is the Heisenberg equation of motion, $\dot{\bm{A}}(t)=i\mathcal{L}\bm{A}(t)$, where $\bm{A}(t)$ denotes a vector of macroscopic variables and $i\mathcal{L}$ is the Liouville operator.
To derive a coarse-grained equation of the density fluctuation, MCT employs the Mori projection operator \cite{Mori1965}.
This formalism derives an equation that contains a memory function as a form of the time-convolution integral.
On the other hand, TMCT employs the Tokuyama--Mori projection operator \cite{Tokuyama1975, Tokuyama1976I}, where the derived equation contains the memory function as a form of the time-convolutionless integral.
The hypothesis concerning the memory function of TMCT is the same as that of MCT, and consequently the memory functions of MCT and TMCT have the same form.
TMCT thus can be studied by the theoretical framework of MCT \cite{Tokuyama2014, Tokuyama2015, Gotze2015}.

%%%%%%%%%%%%%%%%%%%%%%%%%%%%%
%\onecolumngrid
%
\begin{figure*}[t]
  \begin{center}
    \begin{tabular}{l}

      \begin{minipage}{0.795\hsize}
      	\includegraphics[width=0.98\hsize]{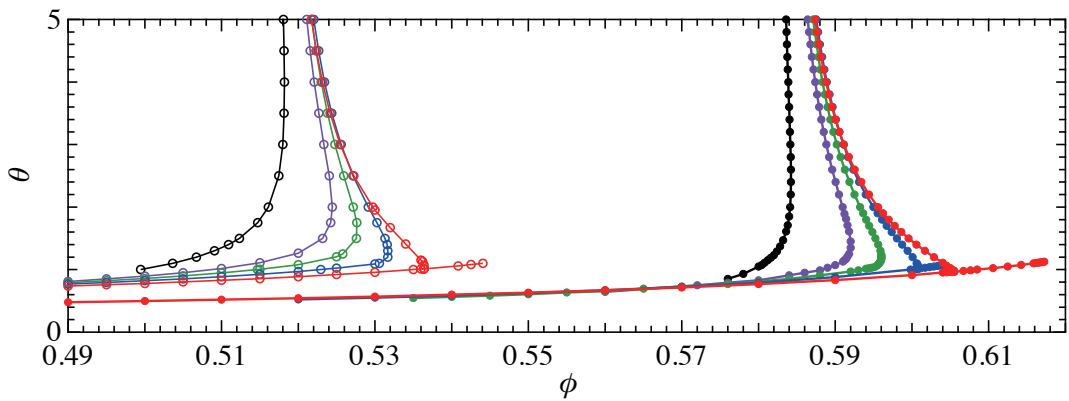}
      \end{minipage}

      \begin{minipage}{0.205\hsize}
		\caption{
			Comparison between the MCT (open circle) and TMCT (filled circle) results of the transition lines of SWS based on PYA: $\varepsilon=0.03$ (red), $0.04$ (blue), $0.05$ (green), $0.06$ (purple), and $0.09$ (black), from right to left.
			The MCT results are identical to those in Ref.~\cite{Dawson2000}.
			\label{fig:liquidGlassTransitionTMCT}
		}
      \end{minipage}
  	\end{tabular}
  \end{center}
\end{figure*}
%\newpage
%\twocolumngrid
%%%%%%%%%%%%%%%%%%%%%%%%%%%%%

TMCT predicts some different features from MCT. 
For example, the initial value of the non-gaussian parameter is a non-zero value in MCT, but $0$ in TMCT \cite{Tokuyama2014}.
In addition, TMCT improves the quantitative features.
For the monodisperse hard-sphere system, Kimura and Tokuyama have solved the TMCT equation by using the static structure factor under the Percus--Yevick approximation (PYA) \cite{Kimura2016}.
The solution has predicted the critical volume fraction $\phi_{\text{c}}=0.582$, while the MCT solution leads to $\phi_{\text{c}}=0.516$ \cite{Bengtzelius1984}.
In this paper, we thus show not only how TMCT qualitatively recovers the MCT predictions for short-range attractive colloids but also how the critical values are quantitatively improved.

The present paper is organized as follows.
Section \ref{sec:method} explains the model we study and the numerical schemes.
Section \ref{sec:result} presents and discusses the kinetic phase diagram obtained numerically.
To support validity of the results, in Sec.~\ref{sec:schematicModel}, a schematic model is investigated analytically.
Section \ref{sec:summary} summarizes this paper.
The details of the analysis for the schematic model are mentioned in the Appendix.

%%%%%%%%%%%%%%%%%%%%%%%%%%%%%
\section{Method} \label{sec:method}
The square-well system (SWS) has been studied as a simple model of short-range attractive colloids \cite{Baxter1968_JCP, Liu1996, Bergenholtz1999_PRE, Fabbian1999, *Fabbian1999_erratum, Foffi2000, Dawson2000, Zaccarelli2001, Gotze2002, Gotze2003, Sperl2004}.
The pairwise potential of SWS is described as $U(r) = \infty~(0<r<d), -u_0~(d<r<d+\Delta), 0~(d+\Delta<r)$, where $d$ denotes the hard-core diameter, $u_0$ the depth of the potential well, and $\Delta$ the width of the attraction.
The equilibrium states are specified by three control parameters: the width parameter $\varepsilon = \Delta / (d + \Delta)$ \cite{Note1}, the volume fraction of the hard spheres $\phi = \pi \rho d^3 / 6$, and the dimensionless temperature $\theta = k_{\text{B}} T/u_0$, where $\rho$ denotes the number density. %cf p.257 in Gotze2009
The molecular dynamics (MD) simulations of SWS have been done for the one-component system \cite{Foffi2002} and binary systems \cite{Sciortino2001, Zaccarelli2002}.

Similarly to the MCT equation for the correlation function of the mode $\rho_{\bm{q}}(t)$ of the density fluctuation, the TMCT equation is solved numerically by using the static structure factor $S_q = \left<\left|\rho_{\bm{q}}(t)\right|^2\right>$ as the initial condition, where the brackets denote an average over an equilibrium ensemble.
The nonergodic transition is intuitively quantified by the Debye--Waller factor $f_q$, which is the long-time limit of the intermediate scattering function $F_q(t)=\left<\rho_{\bm{q}}(t)\rho_{\bm{q}}^*(0)\right>$, i.e., $f_q = \lim_{t\to\infty}F_q(t)/S_q$.
For both MCT and TMCT, the memory function $\mathcal{F}_q$ at the long-time limit is described as
\begin{equation}
	\mathcal{F}_q =\frac{1}{32\pi^2 \rho} \int_{0}^{\infty}dk \int^{\prime} dp \frac{kp}{q^5} S_q S_k S_p v(q,k,p)^2 f_k f_p,
\end{equation}
where the prime at the $p$-integral means that the integration range is restricted to $|q-k|\le p \le q+k$, and $v(q,k,p) = (q^2+k^2-p^2) \rho c_k + (q^2-k^2+p^2) \rho c_p$ with the direct correlation function $c_q = (S_q - 1)/(\rho S_q)$.
The functional $\mathcal{F}_q$ of $f_q$ is called the mode-coupling polynomial which is a central concept of the MCT framework.

The Debye--Waller factor obeys the fixed-point equation $f_q = \mathcal{T}(f_q)$ with
\begin{equation}
	\mathcal{T}(f_q) = \left\{\begin{array}{cl}
		{\displaystyle \frac{1}{1+1/\mathcal{F}_q}} & \text{[MCT]}, \\
		& \\
		{\displaystyle \exp\left(-\frac{1}{\mathcal{F}_q} \right)} & \text{[TMCT]}. \\
	\end{array}\right.	
    \label{eq:fixedEq}
\end{equation}
An ordinary scheme was employed to obtain $f_q$ numerically \cite{Franosch1997}.
The static structure factor of SWS was numerically obtained under PYA \cite{Dawson2000}.
The wavenumber integrals were discretized to $M=500$ points spaced equally, and the cutoff wavenumber was set as $q_{\text{cut}} = 200/d$.
The cutoff was equalized to the previous study for MCT \cite{Dawson2000, Gotze2009}.
Note that we carried out the numerical calculations with $q_{\text{cut}} = 400/d$ to guarantee the independence of the transition points from $q_{\text{cut}}$.

\section{Results and discussion} \label{sec:result}

\subsection{glass-liquid-glass reentrant}

The numerical solution of TMCT describes the liquid-glass-liquid reentrant at small $\varepsilon$. 
Figure \ref{fig:liquidGlassTransitionTMCT} shows the lines connecting the transition points of each $\varepsilon$.
Each transition point was characterized by the maximum eigenvalue $E$, where the bifurcation occurs at which $E=1$ \cite{Note2}.
The liquid-glass transition of TMCT appears at higher volume fractions compared to the MCT results.
The volume fraction of the high temperature limit slightly exceeds the value $\phi_{\text{c}} =0.582$ for the monodisperse hard spheres \cite{Kimura2016} because of the attractive interaction \cite{Dawson2000}.
The shapes of line are qualitatively similar to those of MCT; they are swollen rightward around $\theta \simeq 1$ for small $\varepsilon$.
This indicates the glass-liquid-glass reentry with a decrease of temperature.

%%%%%%%%%%%%%%%%%%%%%%%%%%%%%
\begin{figure*}
	\includegraphics[width=0.98\linewidth]{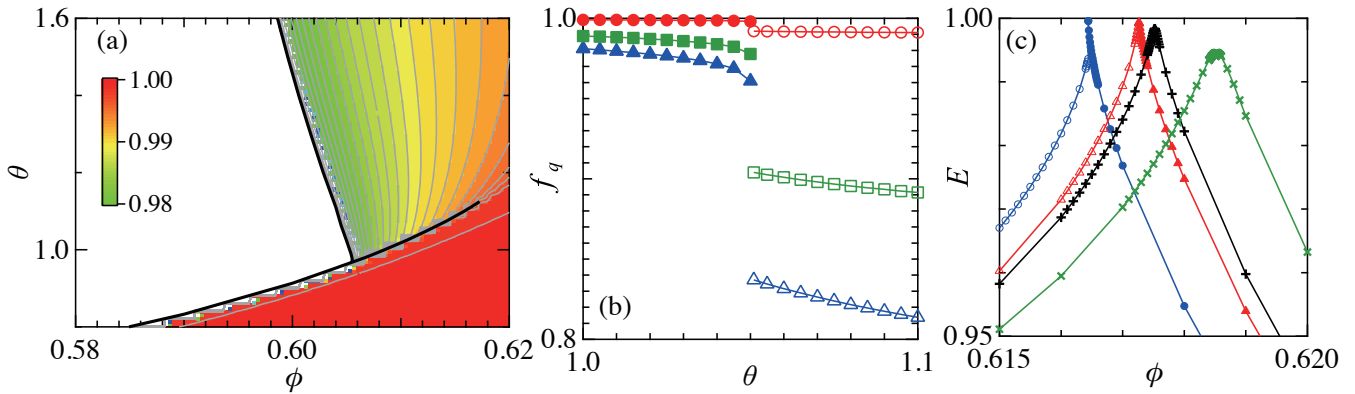}
		\caption{
				Numerical results at $\varepsilon=0.03$ obtained in the TMCT analysis.
				(a) The contour map of the Debye--Waller factor $f_q$ of $q=7.4/d$.
					The white region corresponds to ergodic state, $f_q = 0$.
					The black bold lines indicate the transition line, and the gray ones are the contour line per $0.001$.
					Although the contours near the transition line change in a staircase pattern, they have no physical meanings; the pseudo form is wrongly generated from a software for numerical analysis.
				(b) The temperature dependence of $f_q$ at $\phi=0.61247$.
					The circle (red) indicates result for $q=7.4/d$, the square (green) for $q=10.6/d$, and the triangle (blue) for $q=3.4/d$.
				(c) The $\phi$ dependence of the maximum eigenvalue $E$: $\theta=1.110$ (blue circle), $1.125$ (red triangle), $1.130$ (black plus mark), and $1.150$ (green cross mark). 
				In (b) and (c), the results of the attractive glass is represented by filled symbols and those of the repulsive glass is done by open symbols.
				\label{fig:dwf_and_maxEV}
		}
\end{figure*}
%%%%%%%%%%%%%%%%%%%%%%%%%%%%%

\subsection{glass-glass transition}
At $\varepsilon=0.03$ and $0.04$ in Fig.~\ref{fig:liquidGlassTransitionTMCT}, the TMCT lines corresponding to the attractive glass transition penetrate into the glassy state.
To clarify whether the bifurcation in the glassy state is the glass-glass transition or not, we next focus on the peak value of $f_q$.
Figure \ref{fig:dwf_and_maxEV} (a) illustrates the contour map of the peak value at $\varepsilon=0.03$.
The peak of $f_q$ appears around $q=7.4/d$, which corresponds to the wavenumber where $S_q$ has a peak.
The directions of the contour lines are distinguished with respect to each area of the repulsive and attractive glasses.
Although the peak value continuously changes almost everywhere, it discontinuously changes on the bifurcation line in the glassy state.
Figure \ref{fig:dwf_and_maxEV} (b) shows the value of $f_q$ for three wavenumbers at $\varepsilon=0.03$.
The volume fraction was selected at $\phi=0.61247$ as the bifurcation occurs at $\theta_{\text{c}} = 1.05$.
The behavior of $f_q$ near the glass-glass transition point is the same as that of MCT \cite{Dawson2000}.
In $\theta < \theta_{\text{c}}$, $|f_q - f_{\text{c},q}|$ asymptotically holds the square root variation of $|\theta - \theta_{\text{c}}|$, where $f_{\text{c},q} = \lim_{\theta \nearrow \theta_{\text{c}}} f_{q}$.
This means that the attractive glass appears/disappears as a fold bifurcation.
%In $\theta > \theta_{\text{c}}$, on the other hand, it does not exhibit any power-law behavior.
%; suggesting that the attractive glass is more dominant than the repulsive glass.

\subsection{higher-order singularities}
In this subsection, we confirm that the glass-glass transition line of TMCT ends as well as that of MCT.
Figure \ref{fig:dwf_and_maxEV} (c) shows the $\phi$-dependence of the maximum eigenvalue $E$ for several $\theta$ at $\varepsilon=0.03$.
It clearly shows that there is a marginal temperature $\theta_*$ such that the value of $E$ reaches the unity in $\theta < \theta_*$ and it does not in $\theta > \theta_*$ with controlling $\phi$.
At $\theta = \theta_*$, the eigenvalues of both the repulsive and attractive glasses reach the unity.
It is thus concluded that the glass-glass transition line of $\varepsilon=0.03$ terminates at $(\phi_*, \theta_*) \simeq (0.6173, 1.125)$.
This point has been characterized as $A_3$ singularity (equivalently, cusp bifurcation), which is a higher-order singularity \cite{Fabbian1999, *Fabbian1999_erratum, Bergenholtz1999_PRE, Dawson2000, Zaccarelli2001, Gotze2002, Gotze2003, Sperl2004}.
In this context, the nonergodic transition is classified as the $A_2$ singularity.
Chen et al.~have experimentally proved the existence of the $A_3$ singularity \cite{Chen2003_science}.
Note that the $A_3$ singularity of $\varepsilon=0.04$ is at $(\phi_*, \theta_*) \simeq (0.6039, 1.073)$.
With an increase of $\varepsilon$, the glass-glass transition line disappears at a certain point.
This parameter set is called $A_4$ singularity (equivalently, swallow-tail bifurcation) point \cite{Dawson2000, Gotze2002, Gotze2003, Sperl2004}. 
The TMCT value of $\varepsilon$ at the $A_4$ singularity point is around $0.05$.
As the MCT value is around $0.04$ \cite{Dawson2000}, TMCT extends the $\varepsilon$ range within which the glass-glass transition occurs.
%$\varepsilon_{**}$の増大は何を意味するんだろう？？

%%%%%%%%%%%%%%%%%%%%%%%%%%%%%
\begin{figure}
	\includegraphics[width=0.98\linewidth]{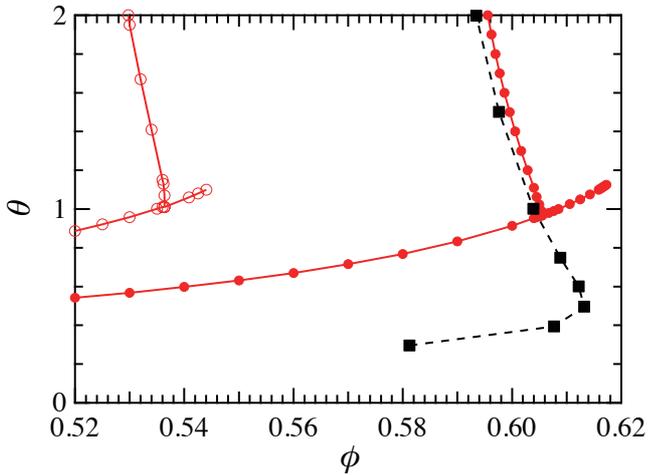}
	\caption{
	The transition lines at $\varepsilon=0.03$.
	The line with filled circles (red) indicates the TMCT result and the line with open circles (red) the MCT result.
	The broken line with squares indicates the MD result of the iso-diffusivity $\tilde{D} = 5\times 10^{-6}$ for A particle of the binary SWS \cite{Zaccarelli2002}.
	\label{fig:comparison}
	}
\end{figure}
%%%%%%%%%%%%%%%%%%%%%%%%%%%%%

\subsection{quantitative comparison of transition points}
We finally compare TMCT with MCT in a quantitative manner.
Figure \ref{fig:comparison} shows the kinetic phase diagram at $\varepsilon=0.03$, in which the TMCT critical values for the one-component SWS are compared with those of MCT and also the MD results for the binary SWS ($\text{A}:\text{B}=50:50$) \cite{Zaccarelli2002}. 
The transition line of the MD simulation was determined by the contour of the normalized diffusivity $\tilde{D} = 5 \times 10^{-6}$ of the A particle, where $\tilde{D}=D/D_0$, $D_0=d_{\text{A}}\sqrt{k_{\text{B}}T/m}$, $D$ denotes the long-time self-diffusion coefficient, and $d_{\text{A}}$ the diameter of the A particle.
The long-time self-diffusion coefficient is an appropriate physical value for a unified comparison between different systems \cite{Narumi2011_MCT4BMLJ}.
The value $5 \times 10^{-6}$ was chosen for the iso-diffusivity line in the high $T$ limit to approach $\phi \simeq 0.58$ \cite{Zaccarelli2002}. 
%, which is based on the fact that the glass transition point obtained experimentally is estimated as $\phi_{\text{g}} = 0.578$ ($\pm 0.7\%$) of polydisperse colloids \cite{vanMegen1995}.
The iso-diffusive line is much closer to the kinetic glass transition line of TMCT without any scaling.
Although the critical temperatures of TMCT overestimate the MD results, we do not judge whether TMCT fails to predict the critical temperature or not. 
Approximation methods (e.g., PYA) for $S_q$ affect the temperature dependence.
A characteristic $T$ of SWS based on the mean-spherical approximation (MSA) is about five times smaller than that based on PYA, while characteristic $\phi$ and $\varepsilon$ are comparable between PYA and MSA \cite{Dawson2000}. 
In fact, the transition line of the TMCT analysis for one-component SWS based on MSA underestimates the MD result.
In addition, the difference might originate from the fact that the simulation was done for binary SWS, while TMCT was applied for one-component SWS.

\section{schematic model} \label{sec:schematicModel}

%%%%%%%%%%%%%%%%%%%%%%%%%%%%%
\begin{figure}
	\includegraphics[width=0.9\linewidth]{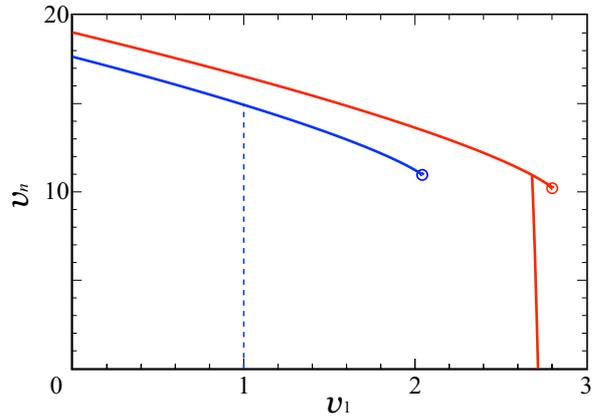}
	\caption{
	The bifurcation diagram on the $v_1$--$v_n$ space of the modified schematic model with $w=1/6$.
	The blue lines correspond to MCT and the red ones do to TMCT.
	The solid lines represent the discontinuous bifurcation and the open circles mark the $A_3$ singularities.
	The dashed line represents the continuous bifurcation, in which the Debye--Waller factor continuously changes across the bifurcation line \cite{Gotze1988}.
	\label{fig:diagramModifiedSM}
	}
\end{figure}
%%%%%%%%%%%%%%%%%%%%%%%%%%%%%

Our numerical results in SWS have shown that TMCT leads to the glass-liquid-glass reentrant, the glass-glass transition, and the higher-order singularities.
However, in a schematic model where MCT predicts both the liquid-glass and glass-glass transition with the $A_3$ singularity \cite{Gotze1988}, G{\"{o}}tze and Schilling have shown that, although the liquid-glass transition occurs in TMCT, the glass-glass transition does not \cite{Gotze2015}.
In this section, we analyze a modified version of the schematic model to support validity of our numerical results.

The model analyzed by G{\"{o}}tze and Schilling assumes a mode-coupling polynomial of a single wavenumber (i.e., $M=1$) as $\mathcal{F} = v_1 f + v_3 f^3$, where $f$ denotes the Debye--Waller factor of $M=1$ and positive coefficients $v_1$ and $v_3$ correspond to control parameters.
We here consider the modified schematic model in which a mode-coupling polynomial is defined by
\begin{equation}
	\mathcal{F} = v_1 f + v_n f^{1+\frac1w},
	\label{eq:modifiedSchematicModel}		
\end{equation}
where, in addition to positive coefficients $v_1$ and $v_n$, a positive coefficient $w$ in the power is another control parameter.

In the modified schematic model \eqref{eq:modifiedSchematicModel}, TMCT predicts the liquid-glass transition, the glass-glass transition, and the higher-order singularities, where the details of the analysis is summarized in the Appendix.
The kinetic phase diagram of the model with $w = 1/6$ (i.e., the power is $7$ in the nonlinear term) includes the $A_3$ singularity as shown in Fig.~\ref{fig:diagramModifiedSM}.
The range of $w$ where the $A_3$ singularity emerges is limited to $w < w_*$ with $w_* = (\sqrt{2}-1)/2 \simeq 0.207$, meaning that the $A_4$ singularity exists at $w=w_*$. 
In the MCT analysis to the modified schematic model, the discontinuous bifurcation line of $w < 1$ terminates and that of $w > 1$ does not.
It implies that the $A_4$ singularity occurs $w_* = 1$ in the MCT analysis.
This difference of $w_*$ between MCT and TMCT is a probable reason why higher-order singularities do not appear in the TMCT analysis by G{\"{o}}tze and Schilling.

The modified schematic model with MCT does not correspond to the short-range attractive colloids because one of the bifurcation line predicted by MCT indicates the continuous bifurcation.
On the other hand, TMCT does not describe any continuous bifurcations \cite{Gotze2015}.
Thus, identifying $(v_1, v_n, w)$ with $(\phi, 1/T, \varepsilon)$, except for the glass-liquid-glass reentrant, the modified schematic model with TMCT qualitatively corresponds to the short-range attraction colloids well.
Nevertheless, it should be noted that the model is just schematic; there is no knowing whether a control parameter such as $\varepsilon$ can shift the nonlinear power.
As the modified schematic model is a single-wavenumber model ($M=1$), it can be interpreted as renormalization from the whole rage of wavenumber alters the power (i.e., $w$) of the nonlinear term.
On the basis of this idea, Tokuyama has described simulation results well by TMCT \cite{Tokuyama2017}.

A similar form of the phase diagram shown in Fig.~\ref{fig:diagramModifiedSM} has been reported by G{\"{o}}tze and Sperl \cite{Gotze2002}.
They have studied a two-wavenumber model with three control parameters:
$\mathcal{F}_1(x_1, x_2) = v_1 x_1^2 + v_2 x_2^2$ and $\mathcal{F}_2(x_1, x_2) = v_3 x_1 x_2$.
In their model, there is a marginal value $v_3^*$ such that the $A_3$ singularity exists in $v_3 > v_3^*$ and it does not in $v_3 < v_3^*$.
At $v_3 = v_3^*$, the discontinuous bifurcation lines collapse.
%Gotze-SperlのモデルをTMCTで解析し，高次特異点が出るのであれば，$\varepsilon_*$について$v_3^*$と対応させて何か理論的に言えるかも．
Note that there are no single-wavenumber models in which the $A_4$ singularity predicted by MCT exists by collapsing the discontinuous bifurcation lines.

\section{Summary} \label{sec:summary}
For SWS as a model of the attractive colloids, we have presented numerical evidence for the existence of the glass-liquid-glass reentrant, the glass-glass transition, and the higher-order (i.e., $A_3$ and $A_4$) singularities in the TMCT analysis.
Compared with the results of MD simulation for a binary colloidal system with short-range attraction, we have clarified quantitative improvement of the critical volume fractions.
As TMCT has the same form of the memory function of MCT, our analysis enhances the utility of the MCT framework for the study of glass transition.
By contrast to the success in the critical volume fraction, the difference of the critical temperatures between theoretical calculations and MD simulations should be addressed.
The TMCT analysis by using the static structure factor obtained from the MD simulation will enable the detailed comparison of the critical temperature with that of the simulation result.

We have analytically studied the modified schematic model defined in Eq.~\eqref{eq:modifiedSchematicModel} to demonstrate that TMCT predicts the higher-order singularities within schematic models.
Except for the glass-liquid-glass reentrant, the modified schematic model qualitatively describes the kinetic phase diagram of the short-range attractive colloids well.
Recalling that $w$ is introduced in the power of the nonlinear term, TMCT suggests that the $A_3$ singularity emerges in the nonlinear power more than $3+2\sqrt{2} \simeq 5.82$, while MCT predicts it in the power more than $2$.
This insensitivity for nonlinearity might cause the quantitative improvement in TMCT.

This paper has concentrated on the static feature described from TMCT; a next issue to consider is the dynamics. 
It is interesting to study typical topics such as the scaling law with regard to the exponent parameter, the logarithmic decay, and stretching features of attractive and repulsive glasses.
Very recently, Tokuyama has shown that the dynamic features obtained in simulations are well-described by TMCT \cite{Tokuyama2017}.
However, we should mention that further investigation is necessary because those works were done approximately by employing a phenomenological approach based on a simplified MCT model \cite{Bengtzelius1984}.
In contrast, our approach can obtain numerical solutions of the TMCT equation without any approximations.
Such dynamic features promote better understanding of the glass transition.
This will be discussed elsewhere.

\begin{acknowledgments}
	We are deeply grateful to Professor Sow-Hsin Chen for suggesting us to apply TMCT for short-range attractive colloidal systems.
	We also wish to thank Dr.~Yuto Kimura for fruitful discussions on numerical calculations.
	This work was partially supported by JSPS KAKENHI Grant No.~JP26400180.
\end{acknowledgments}

\appendix

\section*{APPENDIX} \label{sec:appendix}

The appendix presents details of analysis for the modified schematic model \eqref{eq:modifiedSchematicModel}.

For MCT and TMCT, the fixed-point equation $f=\mathcal{T}(f)$ leads to
\begin{equation}
	v_1 + v_n f^{\frac1w} = \frac{1}{fs},
	\label{eq:G1}
\end{equation}
where $s$ is defined as $1/\mathcal{F}$:

\begin{equation}
	s = \left\{\begin{array}{ll}
		{\displaystyle\frac{f}{1-f}} & \text{[MCT]}, \\
		{\displaystyle\frac{1}{\ln(1/f)}} & \text{[TMCT]}. 
	\end{array}\right.
\end{equation}
%%
%\begin{equation}
%	f = \left\{\begin{array}{ll}
%		{\displaystyle\frac{1}{1+s}} & \text{[MCT]}, \\
%		& \\
%		\exp\left(-s\right) & \text{[TMCT]},
%	\end{array}\right.	
%\end{equation}
%%
Further, the stability matrix $\mathsf{A}$ on the bifurcation points must be unity, that is, 
\begin{equation}
	v_1^c + \left(1+\frac1w \right) v_n^c (f^{c})^{\frac1w} = \frac{1}{(f^c)^{1+\delta_{\text{MCT}}}(s^c)^2},
	\label{eq:G2}
\end{equation}
where the superscript $c$ indicates their critical values and 
\begin{equation}
	\delta_{\text{MCT}} = \left\{\begin{array}{cl}
		1 & \text{[MCT]}, \\
		0 & \text{[TMCT]}. \\
	\end{array}\right.	
\end{equation}
Equations \eqref{eq:G1} and \eqref{eq:G2} reduce to the parametric representation of $v_1^c$ and $v_n^c$:
\begin{eqnarray}
	v_1^c & = & w \left[\left(1+\frac1w \right)G_1 - G_2\right], \label{eq:v1c}\\
	v_n^c & = & \frac{w}{(f^c)^{\frac1w}}(G_2-G_1), \label{eq:vnc}
\end{eqnarray}
with
\begin{equation}
	G_1 = \frac{1}{f^c s^c}~,~~~G_2 = \frac{1}{(f^c)^{1+\delta_{\text{MCT}}}(s^c)^2}.
\end{equation}
%
%%
%\begin{eqnarray}
%	G_1 & = & \frac{1}{f^c s^c}, \\
%	G_2 & = &  \frac{1}{(f^c)^{1+\delta_{\text{MCT}}}(s^c)^2}.
%\end{eqnarray}
%%
Figure \ref{fig:diagramModifiedSM} was drawn based on these equations.
The intercepts of the discontinuous bifurcation lines are derived as follows.
When $v_n = 0$, the critical values are $v_1^c = e$ in TMCT, and $v_1^c = 1$ in MCT for $ w > 1$.
On the other hand, when $v_1 = 0$, Eqs.~\eqref{eq:v1c} and \eqref{eq:vnc} lead to
\begin{equation}
	v_n^c = \left\{\begin{array}{cl}
		{\displaystyle \frac{1}{w}\left(1+w\right)^{1+\frac1w}} & \text{[MCT]}, \\
		{\displaystyle \frac{e}{w}\left(1+w\right)} & \text{[TMCT]}. \\
	\end{array}\right.	
	\label{eq:yIntercept}
\end{equation}
These equations indicate that, for an arbitrary $w>0$, $v_n^c$ of TMCT is larger than that of MCT. 
Note that, in TMCT, $s$ at $v_1 = 0$ is represented by the Lambert W-function $W(x)$ \cite{Tokuyama2017}:
\begin{equation}
	s = -\frac{w}{1+w} W\left(-\frac{1+w}{v_n w}\right).
\end{equation}
Since the domain of $W(x)$ is $x\ge -1/e$, the critical value \eqref{eq:yIntercept} of TMCT is again obtained.

If the $A_3$ singularity occurs, then the exponent parameter $\lambda$ is unity, where $\lambda$ determines the critical exponent $a$ and $b$ of so-called $\beta$ process as $\lambda = \Gamma^2(1+b)/\Gamma(1+2b) = \Gamma^2(1-a)/\Gamma(1-2a)$.
For the modified schematic model, $\lambda$ of TMCT is represented as
\begin{equation}
	\lambda = \frac{s^c}{2}\left[2+\frac1w - s^c\left(1+\frac1w\right)\right].
	\label{eq:lambda}
\end{equation}
In case that $w = 1/2$, G{\"{o}}tze and Schilling have shown that the maximum of $\lambda$ is $2/3$.
However, Eq.~\eqref{eq:lambda} proposes that $\lambda$ of TMCT reaches to $1$ when $w \le w_*$ with $w_* = (\sqrt{2}-1)/2$, i.e., TMCT predicts the $A_3$ and $A_4$ singularities within the schematic model.


\end{document}